# Nonreciprocity engineering in magnetostatic spin waves

Praveen Deorani, Jae Hyun Kwon, and Hyunsoo Yang[*]

*Department of Electrical and Computer Engineering, National University of Singapore, 117576, Singapore*

Magnetostatic surface spin waves (MSSW) excited from a coplanar waveguide antenna travel in different directions with different amplitudes. This effect, called nonreciprocity of MSSW, has been investigated by micromagnetic simulations. The ratio of amplitude of two counter propagating spin waves, the nonreciprocity parameter $\kappa$, is obtained for different ferromagnetic materials, such as NiFe (Py), CoFeAl, yttrium iron garnet (YIG), and GaMnAs. A device schematic has been proposed in which $\kappa$ can be tuned to a large value by varying simple geometrical parameters of the device.

[*] Corresponding author. *E-mail address*: eleyang@nus.edu.sg (H. Yang).

## 1. Introduction

Spin waves are eigen-disturbances in magnetic moments propagating within a magnetic material, such as a ferromagnet, ferrimagnet, or antiferromagnet, via exchange or magnetostatic interactions. Based on the directions of the spin wave propagation ($\vec{k}$) relative to the static magnetization ($\vec{M}$), there are three well known modes of magnetostatic spin waves: (1) magnetostatic surface waves (MSSW), (2) backward volume mode (BVM), and (3) forward volume mode (FVM) [1-5]. In MSSW and BVM, both $\vec{k}$ and $\vec{M}$ lie in the film plane. $\vec{k}$ is perpendicular to $\vec{M}$ in MSSW, but parallel in BVM. In FVM, $\vec{k}$ is in the plane of magnetic film, whereas $\vec{M}$ points out of the film plane.

Spin waves are a subject of great interest because of their potential applications in novel information transfer devices [6-10]. Spin waves are also useful in phase matching of spin torque oscillators [11], and in the enhancement of the spin pumping effect [12]. A recent demonstration of interference-mediated modulation of spin waves offers a new method of engineering spin wave intensity for communication and logic [13, 14]. The tunability of the refractive index and frequency of spin waves over a large range offers an opportunity for technological applications of magnonics [15]. Logic gates based on spin waves have been proposed and experimentally demonstrated [16].

The phenomenon of nonreciprocal wave propagation provides an additional means of controlling the flow of signal and power in the fields of microwave, photonics, and the recently growing areas of magnonics. Nonreciprocity in spin waves is quantified by the parameter $\kappa$, defined as the ratio of amplitudes of counter-propagating spin waves. In magnonic circuits, a large value of $\kappa$ is essential for the realization of logic circuits, interconnects, and switches. In YIG, the $\kappa$ is higher compared to conventional metallic ferromagnets such as Py, however, YIG films are not compatible with silicon-based microfabrication technology. In this work, we address the origin of spin wave nonreciprocity in thin films, and evaluate the value of $\kappa$ in different materials, such as NiFe (Py), CoFeAl, yttrium iron garnet (YIG), and GaMnAs by means of micromagnetic simulations. The results show that $\kappa$ decreases as the saturation magnetization of the material increases, thus explaining a higher value of $\kappa$ in YIG as compared to Py. The $\kappa$ is also shown to increase as the applied bias field increases. In addition, a device geometry for engineering a large value of $\kappa$ is proposed.

**2. Simulation methods**

In this study, micromagnetic simulations based on the Object Oriented MicroMagnetic Framework (OOMMF) [17] are used to investigate the nonreciprocity of spin waves. Simulations are done with a 50 nm × 120 $\mu$m × 50 nm cell size on a cuboidal sample of dimensions 300 $\mu$m × 120 $\mu$m × 50 nm. To excite the spin waves, a pulse field is applied to the sample via a waveguide located at the center of the sample as shown in Fig. 1(a). The waveguide has a width of 2 µm and a thickness of 200 nm, and is separated from the sample by a 50 nm thick insulator. The temporal profile of the pulse is a sinc function with a frequency of 100 GHz, and its spatial profile is given by the Karlqvist equations [18]. The reason for using a sinc pulse of 100 GHz is that, in the frequency domain, this pulse has a uniform distribution over 0 – 15 GHz. A bias field ($H_b$) of 100 Oe is applied along the y-direction. The amplitude of the magnetic field pulse is shown in Fig. 1(b). It must be noted that we have used 1D simulations, with only one cell each in the *y*- and *z*-directions. Since the spin waves in our geometry travel in the *x*-direction, 1D simulations are sufficient to capture the phenomena. In order to confirm this, simulations were also carried out with a cell size of 50 nm × 500 nm × 10 nm, and identical results to those for a larger cell (50 nm × 120 μm × 50 nm) were obtained.

## 3. Results and discussion

Figures 2(a–c) show the magnetization oscillation as a function of time for different modes of spin waves in Py, namely MSSW, BVM, and FVM, respectively, monitored at two different locations (± 10 μm away from the center of spin wave excitation source). The material parameters used for Py are as follows: the Gilbert damping constant $\alpha = 0.01$, the saturation magnetization $M_s = 860 \times 10^3$ A/m, and the exchange stiffness $A = 1.3 \times 10^{-11}$ J/m. Figure 2(d) shows the magnetic field-dependent frequencies of the different modes of spin waves. It is worth noting that a phase difference of π is observed in the BVM as shown in Fig. 2(b) [19], and the FVM is excited only for bias fields higher than that required to saturate $\vec{M}$ in the out-of-plane direction of the film in Fig. 2(d). The bias field-dependence of frequency for the MSSW and BVM is very similar as shown in Fig. 2(d). It can be seen from the dispersion relations of MSSW $\omega^2 = \gamma^2 \mu_0^2 \left[ H_{eff} \left( H_{eff} + M_s \right) + M_s^2 \left( 1 - e^{-2kd} \right)/4 \right]$, and that of BVM $\omega^2 = \gamma^2 \mu_0^2 \left[ H_{eff} \left( H_{eff} + M_s \left( \left(1 - e^{-kd}\right)/kd \right) \right) \right]$, that for small *kd*, both of these equations reduce to $\omega^2 \approx \gamma^2 \mu_0^2 \left[ H_{eff} \left( H_{eff} + M_s \right) \right]$. Here, $H_{eff}$ is the effective field experienced by the magnetic moments, which includes the magnetic anisotropy field, exchange field, and external dc field.

From Fig. 2(a–c), it is clear that the amplitude of precession is different for MSSW travelling in the $+\vec{k}$ and $-\vec{k}$ directions. This effect is called nonreciprocity of spin waves [19-21], and is quantified by $\kappa$, which is defined as the ratio of amplitudes of counter propagating spin wave packets. In this study, we have used the amplitudes of spin waves at distances of +10 $\mu$m and –10 $\mu$m from the source for the calculation of $\kappa$. The nonreciprocity is observed only in MSSW, but not in BVM and FVM.

In order to understand the origin of nonreciprocity in MSSW, simulations were done with only the *x* or *z* components of the pulse field, respectively. Figure 1(b) shows the *x* and *z* components of the pulse field (hereafter called the *x*-pulse and *z*-pulse, respectively). The *z*-pulse changes sign across the waveguide, whereas the *x*-pulse has the same sign throughout the sample. Figures 3(a) and 3(b) show the spin waves generated by the *x*-pulse and *z*-pulse at distances +10 $\mu$m (+*k*) and -10 $\mu$m (–*k*), respectively, from the source. It can be seen that at +10 $\mu$m (–10 $\mu$m), the waves generated from the *x*-pulse and *z*-pulse are in phase (out of phase), resulting in constructive (destructive) interference. In a real experiment, both the *x*-pulse and *z*-pulse cannot be isolated, and are always simultaneously present. Therefore, the total magnitude of the spin waves in a real experiment is the combined effect of both components. Because of interference, the total magnitude of spin waves is larger at +10 $\mu$m than that at –10 $\mu$m from the source. In other words, when the bias field is along the *y*-direction, the amplitude of MSSW propagating in the *x*-direction has a larger amplitude that that in the –*x*-direction.

We have examined the sense of rotation of the magnetization for nonreciprocal spin wave propagation. In principle, it should simply follow the Larmor precession equation $d\vec{M}/dt = -\gamma\, \vec{M} \times \vec{H}$, hence the sense of rotation of the magnetization should be the same for the entire sample, since $\vec{H}$ points in the same direction throughout the sample. In order to confirm this, simulations were carried out to detect the trajectory of the magnetization as a function of time. Figure 4(a) shows the trajectory of the joint $M_x$ and $M_z$ components, as viewed from the –y-direction. The rotation sense is seen to be clockwise for both cases monitored at +10 μm and –10 μm from the spin wave source. Similarly, in Fig. 4 (b) the trajectories of the magnetization in the $M_x$-$M_z$ plane captured at 1 ns are shown as a function of x-coordinate, as viewed from the –y-direction. The $M_x$-$M_z$ components traverse a counter-clockwise trajectory, as the distance from the spin wave source increases. It is also clear from Fig. 4(a) and (b) that the amplitude of trajectories is larger in the case of +k spin waves compared to that of –k waves. From these observations, it is evident that the rotation sense of the magnetization has no correlation with the nonreciprocity of spin waves. By inversing the polarity of the excitation voltage, a distinct π phase shift in the MSSW signal was reported previously [20], however, this will not lead to the reversal of spin wave precession direction from counter-clockwise to clockwise. Rather, it only changes the initial movement of the magnetization direction such that $M_z$ changes sign in the time domain. In Fig. 4 (c) we show the magnetization trajectory under a positive or negative rf pulse. The sense of rotation remains the same (clockwise), but the trajectory is observed to be phase-shifted in time by π radians.

The simulations were extended to obtain the nonreciprocity parameter $\kappa$ in four materials, such as Py, CoFe$_2$Al, YIG, and GaMnAs. Py and YIG are the most widely studied materials in the area of magnonics [5, 22-25]. CoFe$_2$Al is a Heusler alloy, which has recently attracted attention in spintronics research due to its low Gilbert damping constant [26]. GaMnAs represents a class of dilute magnetic semiconductors [27-29]. The parameters for these materials that were used in the simulations are shown in Table 1. The resulting MSSW at distances +10 $\mu$m and −10 $\mu$m are shown in Fig. 5(a-d). For $Co_2FeAl$ and GaMnAs, uniaxial anisotropy was included in the simulations. The obtained values of $\kappa$ are 1.36, 1.28, and 1.7 in Py, $CoFe_2Al$, and YIG, respectively. In the case of GaMnAs, $\kappa$ is found to be 1 based on the maximum amplitude of magnetization oscillation. However, it is difficult to determine $\kappa$, as the wave packet is not well defined. Note that the nonreciprocity ratio of MSSW in various magnetic materials is quite small and similar regardless of the differences in material properties. Previously, spin wave attenuation lengths in these materials were found to be of similar order [30].

A simple analytical expression for the spin wave amplitude is given by [21]

$$m_{\pm} \propto \frac{f}{\gamma} \pm \frac{1}{M_s}\left(H_b^2 - \frac{f^2}{\gamma^2}\right) \qquad (1)$$

where $m_{\pm}$ is the spin wave amplitude in the positive (+) and negative (−) direction from the source, $f$ is the frequency of spin wave, $M_s$ is the saturation magnetization of the propagation medium, $H_b$ is the bias field, and $\gamma$ is the gyromagnetic ratio. YIG has a higher nonreciprocity ratio because of its lower saturation magnetization, as can be seen from Eq. (1). Equation (1) also shows that nonreciprocity can be tuned by $H_b$. The effect of $M_s$ and $H_b$ on nonreciprocity ($m_+/m_-$) as calculated from Eq. (1) is shown in Fig. 5(e). The different values of $M_s$ chosen for this calculation can be found in Table 1. Figure 5(f) shows the effect of $H_b$ on nonreciprocity obtained from micromagnetic simulations using Py parameters. This result is in excellent agreement with the one in Fig. 5(e) obtained from Eq. (1). Among the investigated materials, YIG offers an opportunity to engineer a high ratio of nonreciprocity. However, the fabrication of YIG films is not compatible with conventional semiconductor-based technology. In the case of Py, which is compatible with integrated circuit processing, it is practically impossible to use spin wave nonreciprocity for logic and switch applications due to its small value of nonreciprocity.

As discussed earlier, the origin of nonreciprocity is the interference between waves produced by the *x*- and *z*- component of the pulse field. The interference is constructive (destructive) for spin waves travelling in +*k* (–*k*) direction, leading to a difference in amplitudes of counter propagating waves. The difference in amplitudes quantified by $\kappa$ depends on the effectiveness of interference. For example, if the –*k* waves interfere in a completely destructive way, the amplitude of –*k* waves will be almost zero, and a very high value of $\kappa$ can be expected. Figure 3 shows that the wave amplitude produced by the *z*-pulse is ~5 times smaller than that produced by the *x*-pulse, therefore the interference is not completely effective.

In order to obtain a high value of $\kappa$, the magnitude of the *z*-pulse needs to increase with respect to that of the *x*-pulse. In order to achieve this purpose, we propose a device geometry with two waveguides, one on top and the other below the magnetic layer. As shown in Fig. 6(a), the *z*-pulse fields from these two waveguides add up, whereas the *x*-pulse fields interfere destructively. The value of the *x*-pulse and *z*-pulse for each waveguide is given by the Karlqvist equations [18], and their relative magnitude can be tuned by geometrical parameters such as the width and thickness of the waveguides.

To demonstrate this concept, simulations were performed on a device with the geometry shown in Fig. 6(b). The sample size is 300 $\mu$m × 120 $\mu$m × 50 nm, and the cell size is 50 nm × 120 $\mu$m × 50 nm. The bias field is 100 Oe in the *y*-direction, and the material parameters are those of Py. The spin waves are excited by sending a sinc pulse through the waveguides. It is assumed that the pulse current divides in the two waveguides according to their resistances, which are given by their thicknesses. The thicknesses of the waveguides and insulating layers, shown in Fig. 6(b), were chosen such that the $z$-pulse is ~7 times stronger than the $x$-pulse. This is an optimized ratio between the two components of the pulse, which maximizes the nonreciprocity parameter $\kappa$. The spatial profile of the effective pulse field is plotted in Fig. 6(c). The resulting spin waves in this device are shown in Fig. 6(d). The value of $\kappa$ is found to be 7.8, which is much larger than the corresponding value ($\kappa = 1.3$) for single waveguide devices.

In principle, by using destructive interference, it is possible to completely eliminate the spin wave at –10 $\mu$m from the source, and obtain a very high value of $\kappa$. However, as can be seen from Fig. 3, the spin wave shapes from the $x$-pulse and $z$-pulse are slightly different, therefore it is not possible for them to completely annihilate each other. For perfect cancellation by destructive interference, the waves must have the same shape, which can be achieved by exciting spin waves using a sinusoidal signal. Thus, we have carried out simulations with a sinusoidal excitation from the waveguide. The frequency of excitation is 2.8 GHz, which is the resonance frequency for $H_b = 100$ Oe, and $M_s = 1.07$ T. From the result in Fig. 6(e), the value of $\kappa$ is 54, which is much larger than that obtained for a sinc pulse.

## 4. Conclusion

We have investigated the nonreciprocity of MSSW by micromagnetic simulations. It was shown that the origin of nonreciprocity is the interference of spin waves produced from the $x$ and $z$ components of the excitation pulse. The sense of rotation of magnetization was found to have no correlation with the nonreciprocity. The value of the nonreciprocity parameter $\kappa$ was obtained for different materials by simulations, and found to be small for all cases. A scheme was proposed in which the value of $\kappa$ can be tuned by varying the geometrical parameters of the device.

**Acknowledgment**

This work is partially supported by the Singapore Ministry of Education Academic Research Fund Tier 1 (R-263-000-A46-112) and Singapore National Research Foundation under CRP Award No. NRF-CRP 4-2008-06.

**Table 1.** Parameters used for the different materials in the simulations.

| Material | Damping constant | Saturation magnetization (A/m) | Exchange stiffness (J/m) | Magnetocrystalline anisotropy (J/m$^3$) | References |
|---|---|---|---|---|---|
| Py | 0.01 | 860×10$^3$ | 1.3×10$^{-11}$ | - | |
| YIG | 0.000067 | 150×10$^3$ | 4.2×10$^{-12}$ | - | |
| CoFe$_2$Al | 0.001 | 1053×10$^3$ | 1.5×10$^{-11}$ | Uniaxial, -1000 | [26] |
| GaMnAs | 0.028 | 40×10$^3$ | 2.24×10$^{-13}$ | Uniaxial, -4000 | [31-33] |

Figure captions

Fig. 1. (a) The device geometry used in the simulations. (b) The magnetic field distribution due to *rf* current through the waveguide.

Fig. 2. Different modes of spin waves. The amplitude of oscillation as a function of time in magnetostatic surface mode (a), backward volume mode (b), and forward volume mode (c). (d) The frequency of different spin wave modes as a function of magnetic field.

Fig. 3. Origin of nonreciprocity in magnetostatic surface waves. The waves excited by *x*- and *z*-components of the *rf* pulse interfere constructively for the waves travelling in +$k$ direction (a), whereas they interfere destructively for waves travelling in -$k$ direction (b).

Fig. 4, (a) The trajectory of magnetization in the $M_x$-$M_z$ plane as a function of time is clockwise as viewed from the –*y*-direction at +10 $\mu$m and -10 $\mu$m from the spin wave source. (b) The trajectory of magnetization in the $M_x$-$M_z$ plane as a function of *x*-coordinate as viewed from the –*y*-direction. (c) The trajectory of magnetization in the $M_x$-$M_z$ plane as viewed from the –*y*-direction with different polarities of excitation voltage.

Fig. 5. Nonreciprocity in magnetostatic surface waves in Py (a), CoFe$_2$Al (b), YIG (c), and GaMnAs (d) at $H_b$ = 100 Oe. (e) The dependence of $\kappa$ on $H_b$ for different materials from Eq. (1). (f) The dependence of $\kappa$ on $H_b$ for Py obtained from micromagnetic simulations.

Fig. 6. Engineering nonreciprocity in magnetostatic surface waves. (a) Proposed device geometry with two waveguides. (b) The geometrical parameters of the device used for simulations. (c) The spatial distribution of *rf* magnetic field in the device with geometry in (b). The enhancement in nonreciprocity using the proposed device geometry, when spin waves are excited with a sinc pulse (d), and with a sinusoidal excitation (e).

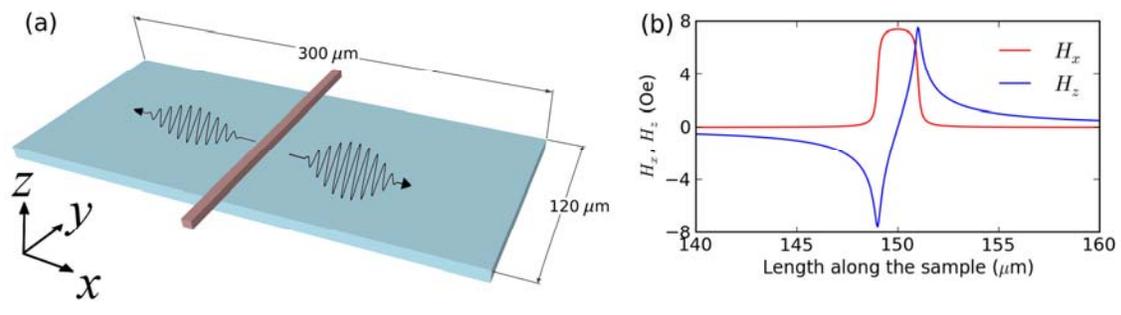

Figure 1

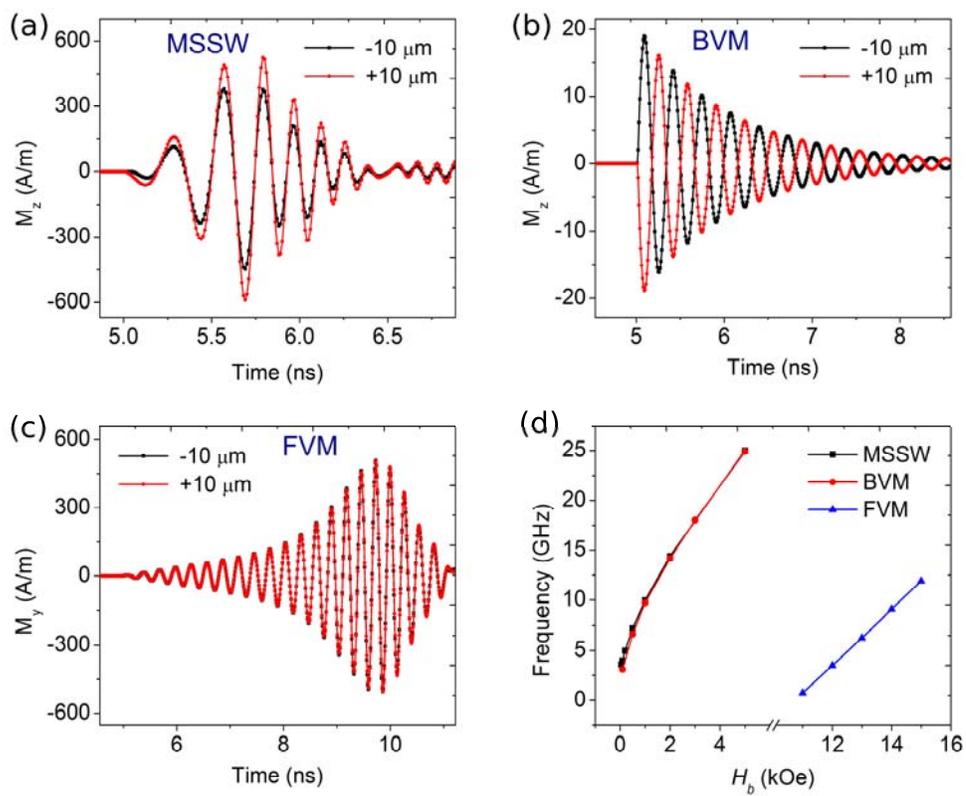

Figure 2

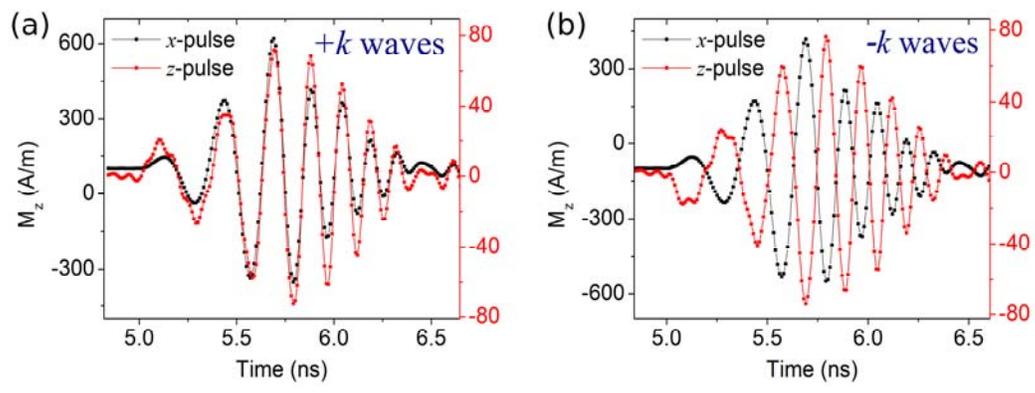

Figure 3

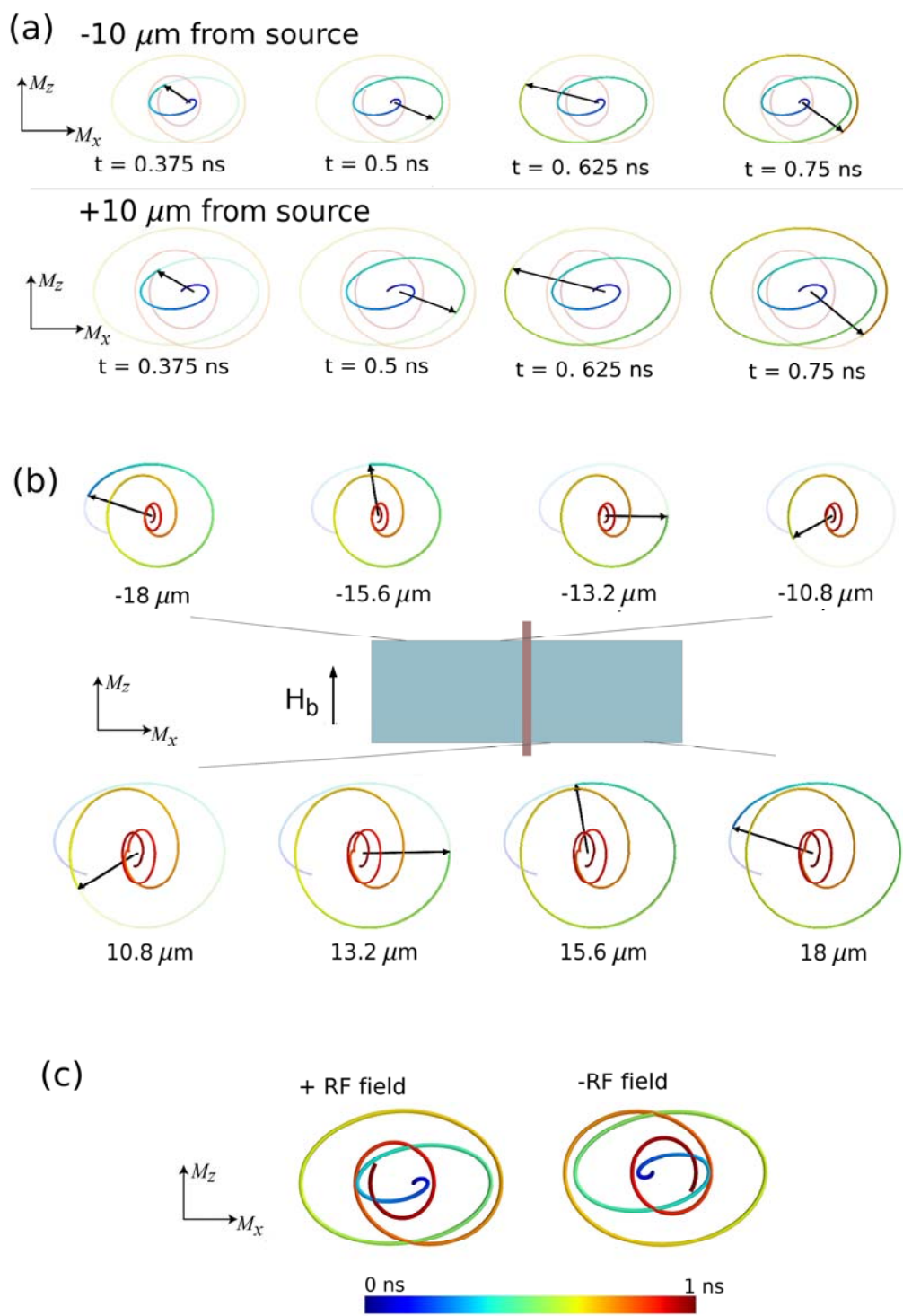

Figure 4

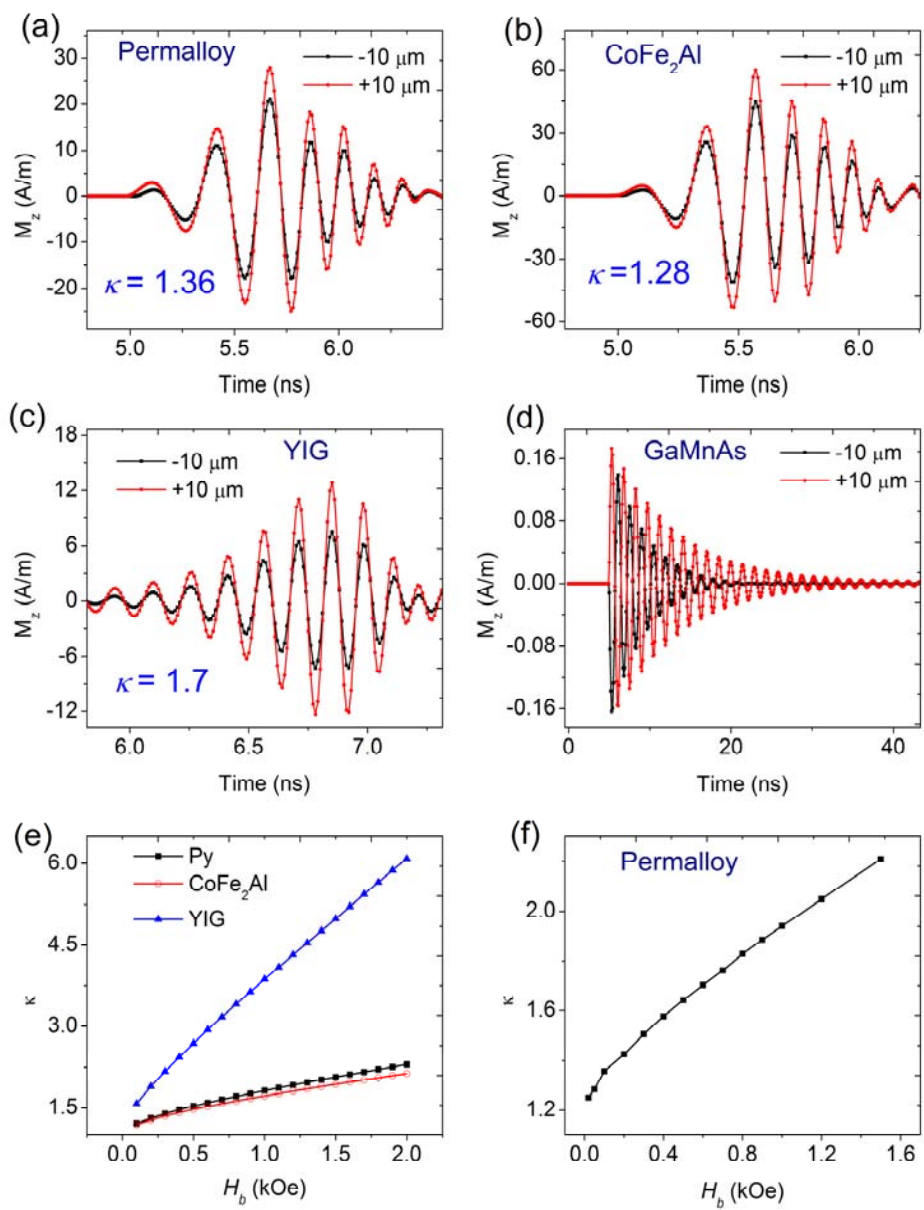

Figure 5

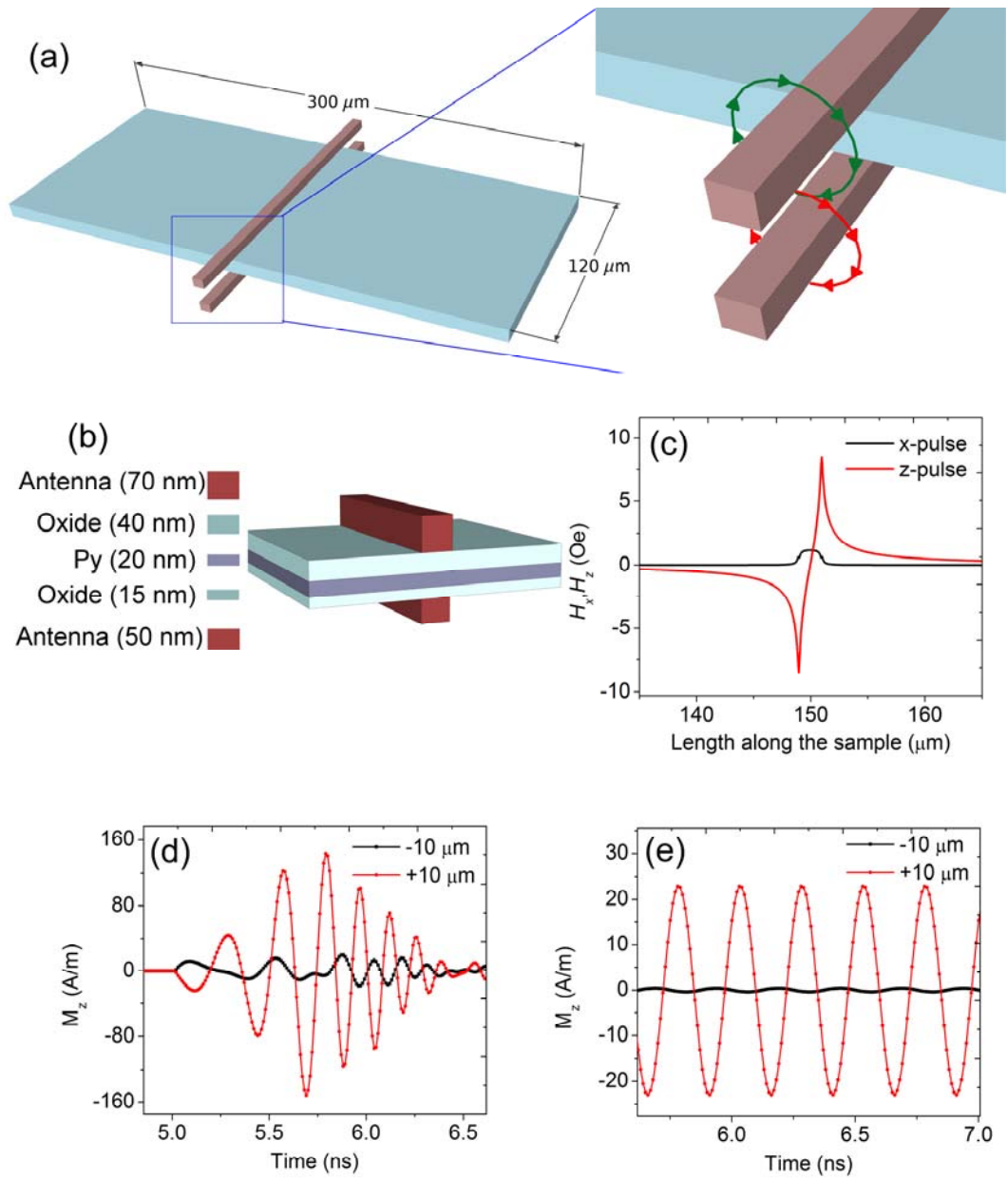

Figure 6